%% file: main.tex
\documentclass[lettersize,journal]{IEEEtran}

\usepackage{amsmath,amsfonts}
\usepackage{algorithmic}
\usepackage{algorithm}
\usepackage{array}
\usepackage[caption=false,font=normalsize,labelfont=sf,textfont=sf]{subfig}
\usepackage{textcomp}
\usepackage{stfloats}
\usepackage{url}
\usepackage{verbatim}
\usepackage{graphicx}
\usepackage{cite}
\usepackage{multirow}
\usepackage{enumitem}
\usepackage[printonlyused]{acronym}
\newacro{ft} [FT] {Fourier transform}
\newacro{mimo} [MIMO] {multiple-input multiple-output}
\newacro{siol} [SIOL] {seamless indoor and outdoor localization}
\newacro{o2i} [O2I] {outdoor-to-indoor}
\newacro{eirp} [EIRP] {equivalent isotropic radiated power}
\newacro{aod} [AoD] {angle-of-departure}
\newacro{aoa} [AoA] {angle-of-arrival}
\newacro{crlb} [CRLB] {Cramér-Rao lower bound}
\newacro{peb} [PEB] {position error bound}
\usepackage{color}

\usepackage[table]{xcolor}
\usepackage{makecell}
\usepackage{hhline}
\newcommand{\vthickline}{\vrule width 0.9pt}
\newcommand{\hthickline}{\noalign{\hrule height 0.9pt}}
\usepackage{tikz} 
\usepackage[utf8]{inputenc}
\usepackage{pgfplots} 
\makeatletter
\newcommand{\gettikzxy}[3]{%
  \tikz@scan@one@point\pgfutil@firstofone#1\relax
  \edef#2{\the\pgf@x}%
  \edef#3{\the\pgf@y}%
}
\usepackage{units} 
\usepackage{bm} 
\usepackage{amssymb} 

\begin{document}
\title{LEO Satellite and RIS: Two Keys to Seamless Indoor and Outdoor Localization}

\author{Pinjun~Zheng, Xing~Liu, Yuchen Zhang, 
Jiguang~He,~\IEEEmembership{Senior Member,~IEEE}, 
Gonzalo~Seco-Granados,~\IEEEmembership{Fellow,~IEEE},
and~Tareq~Y.~Al-Naffouri,~\IEEEmembership{Fellow,~IEEE}

}

\markboth{draft}{draft}
\maketitle

\begin{abstract}
The contemporary landscape of wireless technology underscores the critical role of precise localization services. Traditional global navigation satellite systems (GNSS)-based solutions, however, fall short when it comes to indoor environments, and existing indoor localization techniques such as electromagnetic fingerprinting methods face challenges of additional implementation costs and limited coverage. This article explores an innovative solution that blends low Earth orbit (LEO) satellites with reconfigurable intelligent surfaces (RISs), unlocking its potential for realizing seamless indoor and outdoor localization (SIOL) with global coverage. After a comprehensive review of the distinctive characteristics of LEO satellites and RISs, we showcase their potential for SIOL applications through two case studies on position error bound evaluation. Finally, we discuss system architectures and highlight open challenges in such systems.
\end{abstract}


\section{Introduction}
Nowadays, the significance of localization (or positioning) within modern wireless systems has steadily grown. Releases 18 and 19 of the 3rd Generation Partnership Project (3GPP) have consistently highlighted localization's importance in enhancing communication efficiency, coverage, reliability, and integrity. For a long period, user localization mainly relied on global navigation satellite systems (GNSS), which typically operate at altitudes on the order of \unit[20,000]{km} in the medium Earth orbit (MEO) and are primarily designed for providing positioning, navigation, and timing (PNT) services. Concurrently, cellular radio localization methods have been developed and continuously evolved since the inception of first-generation (1G) cellular technology~\cite{Del2018Survey}.

Despite the global coverage and high performance of GNSS in outdoor environments, indoor users have been unable to access GNSS services, a limitation also observed in cellular-based localization. 
Consequently, the focal point for the location of mobile radio users has shifted from predominantly outdoor to indoor environments~\cite{Del2018Survey}.
Indoor localization leveraging a variety of technologies such as WiFi, Bluetooth, barometers, terrestrial base stations, and more was first highlighted in Release~13 of the 3GPP. Subsequently, Release~14 took indoor localization a step further by emphasizing higher accuracy and shorter service latency~\cite{Lin2022An}.
However, existing cellular-based indoor localization methods such as fingerprinting methods incur additional implementation costs for network operators. Moreover, these methods are usually proprietary, leading to a lack of global coverage.

\begin{figure*}[t]
  \centering
  \includegraphics[width=0.9\linewidth]{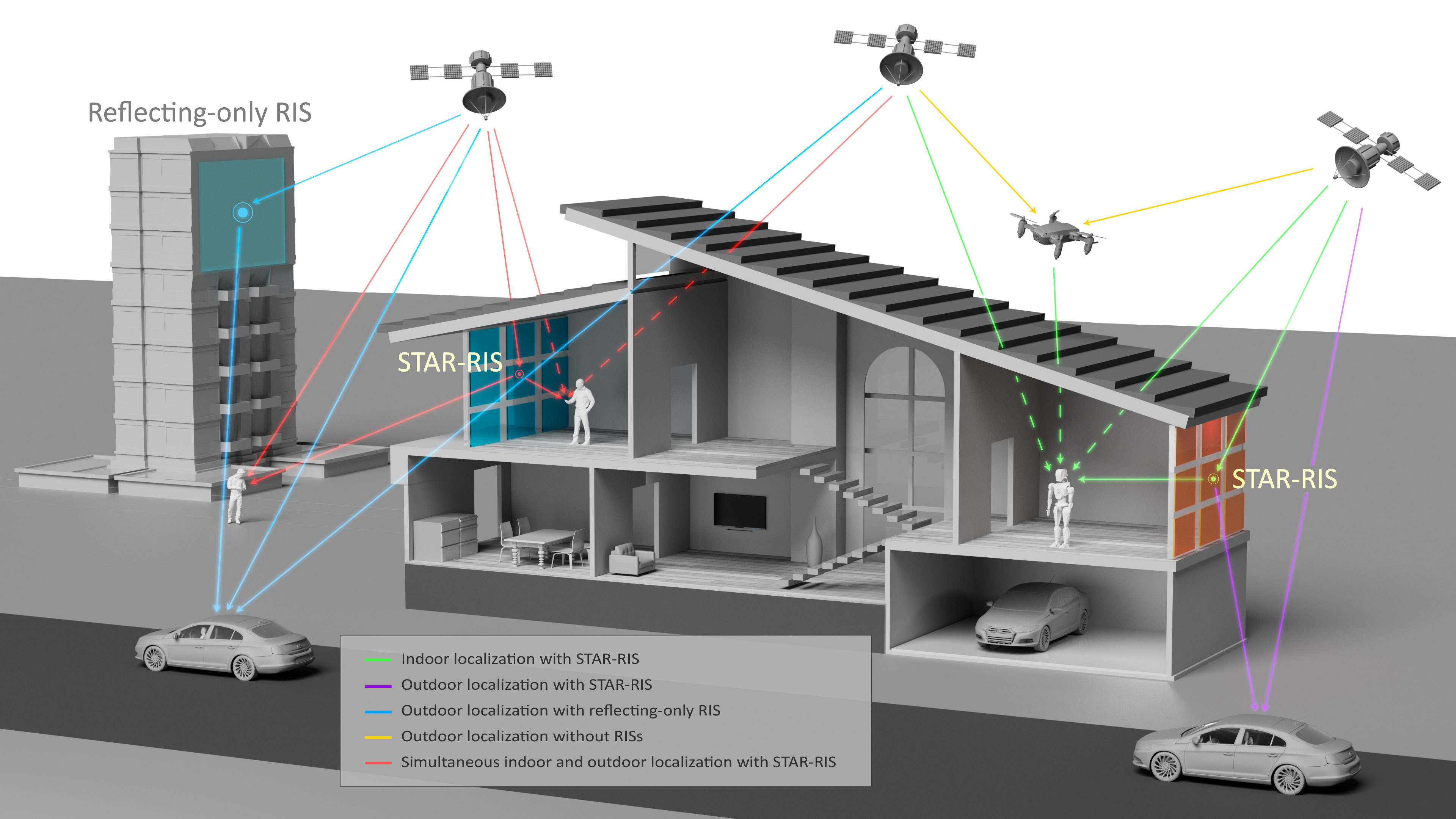}
  \caption{ 
      Conceptual diagram of seamless indoor and outdoor localization services enabled by LEO satellites and RISs. With the involvement of STAR-RISs, LEO satellite signals can extend into indoor areas thus enabling and enhancing indoor localization, while outdoor users can benefit from both reflecting-only RISs and STAR-RISs.
    }
  \label{fig_system}
\end{figure*}

A promising alternative for achieving universal indoor localization is the utilization of low Earth orbit (LEO) satellites. LEO satellites operate at altitudes ranging from a few hundred to \unit[2,000]{km}, enabling faster satellite motion and shorter signal propagation distances compared to MEO and geostationary Earth orbit (GEO) satellites. As a result, LEO satellites yield stronger signal reception capabilities, offering substantial potential for indoor localization. Nonetheless, while LEO satellite signals can be receivable by indoor users, they still cannot deliver the required signal strength for achieving high-accuracy indoor localization. This limitation arises from the fact that LEO satellites face challenges similar to those encountered by GNSS. Namely, satellite signals always undergo significant degradation within urban regions due to the obstructive, attenuation, reflective, and diffractive effects of surrounding buildings~\cite{Zheng2023Attitude}. 
To overcome these inherent challenges, reconfigurable intelligent surfaces (RISs) emerge as a powerful solution~\cite{Wang2024Wideband}. RISs can intelligently manipulate wireless channels, ensuring effective and reliable satellite signal reception for ground users. 
In particular, active RISs can strengthen weak satellite signals, and simultaneously transmitting and reflecting RISs (STAR-RISs) can act as bridges connecting indoor and outdoor environments, further enhancing signal reception and thus improving localization performance.
Therefore, the collaborative utilization of LEO satellites and RISs holds the potential to provide a universal indoor and outdoor localization service. 
Fig.~\ref{fig_system} illustrates a conceptual diagram showcasing the integration of LEO satellites and RISs providing seamless indoor and outdoor localization services.

This article presents a feasibility study exploring the utilization of LEO satellites and RISs for achieving \ac{siol}. We begin by introducing the individual features and recent advances of LEO satellites and RISs. Subsequently, we discuss their potential for a systematic integration. To reveal the potential of the proposed system, several case studies are conducted, which evaluate the \ac{peb} for RIS-aided localization based on LEO satellite signals. Finally, possible system architectures and several open challenges are discussed, followed by the conclusion of this work. 

\section{LEO Satellite}

The concept of LEO satellites is not novel, with the first one launched as early as 1957. By the 1980s, the potential for narrowband communications using LEO satellites was being explored. However, the high costs associated with them prevented their widespread adoption.
With the advancement of launch vehicles, small satellites, and space technologies over the past few decades, the costs of LEO satellites have dropped dramatically. This has driven the recent surge in the popularity and deployment of LEO satellites. Table~\ref{tab1} provides an overview of selected satellite constellations in LEO, MEO, and GEO. A comparison of satellite features across these orbits can be found in Fig.~\ref{fig:spider}, highlighting the benefits of LEO satellites in terms of launch cost, latency, data throughput, and maneuverability. As indicated in Table~\ref{tab1}, LEO constellations are designed to cater to a range of services, including PNT, broadband Internet, voice and data service, internet of things (IoT), machine-to-machine (M2M) communications, and remote sensing.

\begin{table*}[tbp]
\centering
\caption{Comparison of selected satellite constellations across various orbits.}
\setlength{\tabcolsep}{1.mm}
\renewcommand{\arraystretch}{1.3}
\begin{tabular}{!\vthickline c !\vthickline c|c|c|c|c|c!\vthickline}
\hthickline 
\textbf{Regime} & \textbf{Constellation} & \textbf{Orbital height} & \textbf{Quantity} & \textbf{Bands} &\textbf{Modulation} & \textbf{Services}  \\
\hthickline
\multirow{5}{*}{GEO} & \cellcolor{green!5}Intelsat & \cellcolor{green!5}35,786~km & \cellcolor{green!5} 52 & \cellcolor{green!5}C, Ku, Ka & \cellcolor{green!5}$\begin{array}{c} \text{QPSK}\\
\text{8PSK} \\
\text{16APSK \& 32APSK} \\
\text{QAM}\end{array}$ & \cellcolor{green!5} $\begin{array}{c}\text{Broadcasting}\\
\text{Data \& communications}\\
\text{Mobile connectivity} \end{array}$\\\hhline{~|*{6}{-}} 
  &\cellcolor{green!5}Eutelsat  &\cellcolor{green!5} 35,786~km &\cellcolor{green!5} 36 &\cellcolor{green!5}C, Ku, Ka &\cellcolor{green!5}$\begin{array}{c} \text{QPSK}\\
\text{8PSK} \\
\text{16APSK} \end{array}$ &\cellcolor{green!5} $\begin{array}{c} \text{Video \& data services}\\
\text{Mobility} \\
\text{Government Services} \end{array}$\\ \hhline{~|*{6}{-}} 
 
& \cellcolor{green!5} Arabsat & \cellcolor{green!5}35,786~km & \cellcolor{green!5} 8 & \cellcolor{green!5}C, Ku, Ka & \cellcolor{green!5}$\begin{array}{c} \text{QPSK} \\
\text{8PSK} \\
\text{DVB-S2X}\end{array}$ & \cellcolor{green!5} $\begin{array}{c} \text{Broadcasting} \\
\text{Telecommunications} \\
\text{Broadband Internet}\end{array}$ \\
\hthickline 
\multirow{5}{*}{MEO} & \cellcolor{blue!5}GPS & \cellcolor{blue!5}20,180~km & \cellcolor{blue!5}31 & \cellcolor{blue!5}L &\cellcolor{blue!5}BPSK, BOC & \cellcolor{blue!5}PNT \\ \hhline{~|*{6}{-}} 
  & \cellcolor{blue!5}GLONASS & \cellcolor{blue!5}19,130~km & \cellcolor{blue!5}24 & \cellcolor{blue!5}L &\cellcolor{blue!5} BPSK & \cellcolor{blue!5}PNT \\ \hhline{~|*{6}{-}} 
 &\cellcolor{blue!5}Galileo  & \cellcolor{blue!5}23,220~km&\cellcolor{blue!5}30 &\cellcolor{blue!5}L &\cellcolor{blue!5} BOC, CBOC &\cellcolor{blue!5}PNT \\ \hhline{~|*{6}{-}} 
 &\cellcolor{blue!5}BeiDou  & \cellcolor{blue!5}21,530~km&\cellcolor{blue!5}$\begin{array}{c}27 \\
\text{\!(+\! 3\! IGSO\! \&\! 5\! GEO)\!}\end{array}$ &\cellcolor{blue!5}L &\cellcolor{blue!5} QPSK, BOC &\cellcolor{blue!5}PNT \\
\hthickline
\multirow{14}{*}{LEO}  
&\cellcolor{red!5}Xona  &\cellcolor{red!5}525~km &\cellcolor{red!5}258 &\cellcolor{red!5}L, C &\cellcolor{red!5}- & \cellcolor{red!5}PNT\\ \hhline{~|*{6}{-}}
&\cellcolor{red!5}$\begin{array}{c}\text{Starlink Gen.} 1\\
\text{Starlink Gen.} 2\end{array}$ & \cellcolor{red!5}$\begin{array}{c}335.9 \sim 570 \mathrm{~km} \\
340 \sim 614 \mathrm{~km}\end{array}$ & \cellcolor{red!5}$\begin{array}{c}\text{11,926} \\
\text{30,000}\end{array}$ & \cellcolor{red!5} L,\! S,\! Ku,\! Ka,\! V  &\cellcolor{red!5}OFDM & \cellcolor{red!5} $\begin{array}{c}\text{Broadband Internet}\\ \text{\& direct-to-cell services} \end{array}$  \\ \hhline{~|*{6}{-}}
 &\cellcolor{red!5}OneWeb &\cellcolor{red!5}1,200~km & \cellcolor{red!5}648 &\cellcolor{red!5}Ku  &\cellcolor{red!5}OFDM & \cellcolor{red!5}Broadband Internet  \\ \hhline{~|*{6}{-}}
 &\cellcolor{red!5}$\begin{array}{c}\text{Telesat Phase.} 1 \\
\text{Telesat Phase.} 2\end{array}$ & \cellcolor{red!5}$\text{1,015} \sim \text{1,325} \mathrm{~km}$ & \cellcolor{red!5}$\begin{array}{c}298 \\
\text{1,373}\end{array}$ &\cellcolor{red!5}Ku, Ka  &\cellcolor{red!5}- & \cellcolor{red!5}Broadband Internet  \\ \hhline{~|*{6}{-}}
 &\cellcolor{red!5}Hongyan & \cellcolor{red!5}1,100 $\sim$ 1,175 ~km & \cellcolor{red!5}320 & \cellcolor{red!5}Ka &\cellcolor{red!5}- & \cellcolor{red!5}Broadband Internet \\ \hhline{~|*{6}{-}}
 &\cellcolor{red!5}Kuiper & \cellcolor{red!5}$590 \sim 630 \mathrm{~km}$ & \cellcolor{red!5}3,236 &\cellcolor{red!5}Ka  &\cellcolor{red!5}- & \cellcolor{red!5}Broadband Internet  \\ \hhline{~|*{6}{-}}
 &\cellcolor{red!5}Globalstar & \cellcolor{red!5}1,414~km & \cellcolor{red!5}48 &\cellcolor{red!5} L, S &\cellcolor{red!5}OQPSK  & \cellcolor{red!5}Voice and data services  \\ \hhline{~|*{6}{-}}
 &\cellcolor{red!5}$\begin{array}{c}\text{Orbcomm Gen.} 1\\
\text{Orbcomm Gen.} 2\end{array}$ & \cellcolor{red!5}$\begin{array}{c} 720 \mathrm{~km} \\750 \mathrm{~km}\end{array}$ & \cellcolor{red!5}$\begin{array}{c}36 \\18\end{array}$ & \cellcolor{red!5}VHF &\cellcolor{red!5}	SDPSK & \cellcolor{red!5}IoT \& M2M communications \\ \hhline{~|*{6}{-}}
 &\cellcolor{red!5}Iridium NEXT & \cellcolor{red!5}781~km & \cellcolor{red!5}66 & \cellcolor{red!5}L, Ka  &\cellcolor{red!5}QPSK & \cellcolor{red!5}$\begin{array}{c}\text{Voice and data}\\
\text{M2M communications} \end{array}$ \\ \hhline{~|*{6}{-}}
 &\cellcolor{red!5}Dove \& \cellcolor{red!5}SkySat (Planet) &\cellcolor{red!5}$400 \sim 600 \mathrm{~km}$  &\cellcolor{red!5}194  &\cellcolor{red!5}Multispectral  &\cellcolor{red!5}- &\cellcolor{red!5}Earth observation  \\ \hhline{~|*{6}{-}}
 &\cellcolor{red!5}ICEYE & \cellcolor{red!5}560 $\sim$ 580~km  &\cellcolor{red!5}27  &\cellcolor{red!5}X &\cellcolor{red!5}- &\cellcolor{red!5}Earth observation  \\
\hthickline 
\end{tabular}
\label{tab1}
\end{table*} 

\subsection{LEO Satellite-Based Localization}

In recent years, the localization potential of LEO satellites has attracted significant attention, leading to the development of various implement strategies~\cite{9840374}.

\begin{itemize}
    \item \textbf{Dedicated LEO Constellations}: Systems such as Xona Space Systems, TrustPoint, Geely, and Future Navigation are being developed explicitly for PNT services. These dedicated LEO-PNT systems deliver critical data, including ephemeris, clock biases, and drift corrections, to enable precise PNT solutions. Signal structures like GNSS-like signals, orthogonal frequency-division multiplexing (OFDM), and chirp spread spectrum (CSS) are currently under investigation for these applications.
    \item \textbf{Signals of Opportunity (SoOP):} LEO satellites originally designed for non-navigation purposes are now being repurposed for localization through Signals of Opportunity (SoOP). Constellations like Starlink, Orbcomm, Globalstar, and Iridium are notable in this field. Research in this area focuses on advanced signal processing techniques and receiver architectures optimized for opportunistic localization scenarios.
\end{itemize}

\begin{figure}[tbp]
\centering
\hspace{-2em}
\include{figures/Spider_Chart}
\vspace{-30pt}
\caption{Comparative features of satellites in GEO, MEO, and LEO.}
\label{fig:spider}
\end{figure}

LEO satellite signals can be processed to extract various observations, including pseudo-range, carrier phase, Doppler shift, \ac{aoa}, and \ac{aod}, which are then utilized for localization. The requirements and characteristics of these observations are outlined in Table~\ref{tab:leo_pnt_observations}. Depending on the type of observations, different localization algorithms are required to ensure suitability and effectiveness. For example, when using carrier-phase observations to perform snapshot localization, integer least-squares (ILS) optimization is an appropriate approach. However, such a solution necessitates a process known as ambiguity resolution to address unknown integer ambiguities. Besides, the algorithm design also depends on the constraints of unknown variables. For example, when estimating the orientation/attitude (which is constrained in the special orthogonal group in three dimensions SO(3)) in scenarios with antenna arrays on the receiver, constrained optimization techniques, such as Riemannian manifold-based approaches, are effective tools. Furthermore, for dynamic tracking applications, filters such as the Kalman filter and its variants, particle filter, and factor graph, are often adopted to maintain continuous and reliable estimation.

\begin{table*}[tbp]
\centering
\caption{Requirements and Characteristics of LEO Satellite Observations in Localization}
\renewcommand{\arraystretch}{1.3}
\begin{tabular}{!\vthickline>{\centering\bfseries}m{2cm} !\vthickline >{\centering\arraybackslash}m{7cm}|>{\centering\arraybackslash}m{7.5cm} !\vthickline}
\hthickline
\textbf{Observation} & \textbf{Requirements} & \textbf{Characteristics} \\
\hthickline
\cellcolor{cyan!5} Pseudo-range & \cellcolor{cyan!5} \begin{itemize}[leftmargin=*, noitemsep, topsep= 0.5\baselineskip, partopsep=0pt, parsep=0pt, after=\vspace{-0.7\baselineskip}] \item Satellite atomic clocks \item Time synchronization \item Clock bias correction \item Accurate signal travel time estimation \end{itemize} & \cellcolor{cyan!5} \begin{itemize}[leftmargin=*, noitemsep, topsep= 0.5\baselineskip, partopsep=0pt, parsep=0pt, after=\vspace{-0.7\baselineskip}] \item Approximate distance measurement with higher noise \item Prone to clock bias, atmospheric delays, and multipath effects \end{itemize} \\
\hline
\cellcolor{yellow!5}  Carrier-phase & \cellcolor{yellow!5} \begin{itemize}[leftmargin=*, noitemsep, topsep= 0.5\baselineskip, partopsep=0pt, parsep=0pt, after=\vspace{-0.7\baselineskip}] \item High-precision clock stability \item Stable oscillators for frequency and phase stability \item Doppler shift compensation \item Reliable signal tracking (e.g., phase-locked loop) \end{itemize} & \cellcolor{yellow!5} \begin{itemize}[leftmargin=*, noitemsep, topsep= 0.5\baselineskip, partopsep=0pt, parsep=0pt, after=\vspace{-0.7\baselineskip}] \item Millimeter-level precision \item Presence of phase ambiguity \end{itemize} \\
\hline
\cellcolor{cyan!5} Doppler shift & \cellcolor{cyan!5} \begin{itemize}[leftmargin=*, noitemsep, topsep= 0.5\baselineskip, partopsep=0pt, parsep=0pt, after=\vspace{-0.7\baselineskip}] \item Stable oscillators for precise frequency tracking \item Rapid signal acquisition \item Adaptive tracking algorithms (e.g., frequency-locked loop)\end{itemize} & \cellcolor{cyan!5} \begin{itemize}[leftmargin=*, noitemsep, topsep= 0.5\baselineskip, partopsep=0pt, parsep=0pt, after=\vspace{-0.7\baselineskip}] \item Highly sensitive to relative motion \item Exhibits short-term frequency variability \end{itemize} \\
\hline
\cellcolor{yellow!5} AoA/AoD & \cellcolor{yellow!5} \begin{itemize}[leftmargin=*, noitemsep, topsep= 0.5\baselineskip, partopsep=0pt, parsep=0pt, after=\vspace{-0.7\baselineskip}] \item Multi-element antenna arrays \item Element synchronization for phase accuracy \item Advanced signal processing algorithms (e.g., MUSIC, ESPRIT) \end{itemize} & \cellcolor{yellow!5} \begin{itemize}[leftmargin=*, noitemsep, topsep= 0.5\baselineskip, partopsep=0pt, parsep=0pt, after=\vspace{-0.7\baselineskip}] \item[] Resolution and accuracy depend on array geometry (e.g., linear, circular, planar) and the number of elements. More elements generally provide higher resolution at the expense of increased complexity and cost. \end{itemize} \\
\hthickline
\end{tabular}
\label{tab:leo_pnt_observations}
\end{table*}

\subsection{Why Is the LEO Satellite a Key Enabler for SIOL?}

\begin{figure*}[t]
  \centering
  \includegraphics[width=\linewidth]{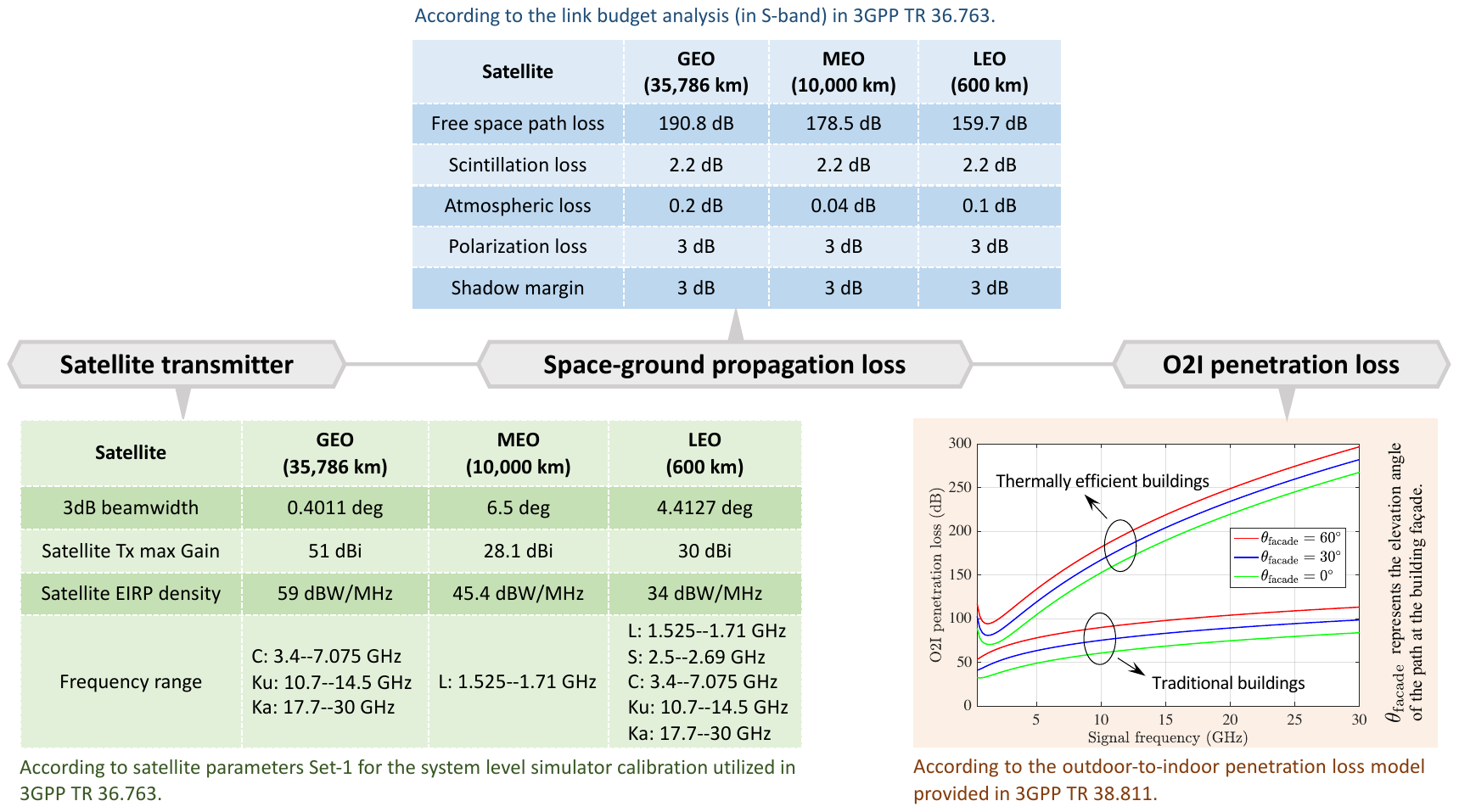}
  \caption{ 
      A summary of satellite transmitter configurations, space-to-ground propagation losses, and outdoor-to-indoor penetration losses for GEO, MEO, and LEO satellites.
    }
  \label{fig_LinkBudget}
\end{figure*}

Despite the strong localization capabilities of LEO satellites, numerous alternative satellite systems are also available for providing localization services. However, we contend that LEO satellites are among the most promising candidates for \ac{siol} applications, for the following reasons:

\begin{itemize}
\item \textbf{Stronger Power Reception:} Based on 3GPP TR 36.763~\cite{3GPP36763} and TR 38.811~\cite{3GPP38811}, Fig.~\ref{fig_LinkBudget} demonstrates a typical setup of transmitter configurations, space-ground propagation losses, and \ac{o2i} penetration losses for GEO, MEO, and LEO satellites. GEO satellites are generally unsuitable for localization applications due to their limited geometric diversity and high latency. The comparison between MEO and LEO satellites shows that, while they have similar antenna gains, the transmit \ac{eirp} density of MEO satellites is typically about 10~dB/MHz higher. However, the closer proximity of LEO satellites to the Earth's surface reduces propagation path loss by approximately 20~dB, resulting in a net 10~dB improvement in received signal strength compared to MEO satellites. This gain is consistent across indoor and outdoor environments, as \ac{o2i} penetration loss is independent of satellite altitude, depending instead on factors such as signal frequency, angle of incidence, and building type, as illustrated in Fig.~\ref{fig_LinkBudget}. The enhanced signal reception of LEO satellites can enable indoor localization and enhance outdoor localization, while also bolstering their resistance to interference, jamming, and spoofing attacks. 

\item \textbf{Better Compatibility with Terrestrial Networks:} LEO satellites can support high-throughput and low-latency requirements, making them a better fit for integration into next-generation networks designed for diverse use cases including localization. In addition, as shown in Table~\ref{tab1}, LEO satellites adopt efficient signal modulations such as OFDM, which align well with those commonly employed in 5G. There are ongoing standardization activities aimed at integrating LEO satellite networks into 5G and beyond~\cite{Darwish2022LEO}. Terrestrial 5G and beyond networks are capable of providing localization services across small areas, covering both indoor and outdoor environments. Integrating these networks with LEO satellite systems holds the potential to deliver uninterrupted, high-performance \ac{siol} everywhere.

\item \textbf{Other Benefits:} Apart from the aforementioned benefits, LEO satellites offer several other advantages that enhance localization capabilities in general: (i) The large LEO constellations (consisting of hundreds or even thousands more satellites than MEO and GEO) can provide superior visibility and improved localization accuracy through multi-satellite cooperation; (ii) The stronger Doppler shifts of LEO satellites yield more location-related information; (iii) The larger bandwidth used in LEO satellite systems (as shown in Fig.~\ref{fig_LinkBudget}) enables higher delay resolution in the received signal.
\end{itemize}

\section{Reconfigurable Intelligent Surface}

Emerging as a transformative technology, RISs have their historical roots dating back to the 1960s, with the precursor known as the reflectarray antenna~\cite{Berry1963Reflectarray}. Over the previous decades, these early studies have paved the way for the development of the modern RIS. Driven by the ever-increasing demand for the intelligent control of radio propagation environments in contemporary wireless systems, the RIS technology has attracted great attention from both industry and academia. The RIS may possess a multitude of structures, e.g., uniform planar array (UPA), uniform circular array (UCA), volumetric array, and conformal arrays. Each RIS radiating element can be digitally controlled and programmed by the usage of positive-intrinsic-negative (PIN) diodes and the change of bias voltage. Besides, there are several technologies to implement RISs, such as varactor diodes, liquid crystal, and leaky-wave antennas. 

\subsection{RIS-Aided Localization}

For localization applications, an RIS can not only act as a new synchronized location reference but also provide additional geometric measurements thanks to its high angular resolution. With RISs being introduced properly, it is possible to perform localization in many previous infeasible scenarios~\cite{Keykhosravi2023Leveraging}. Before revealing the usage of RISs in \ac{siol}, this subsection first briefly recaps the recent advances in the field of RIS-aided localization.

\subsubsection{Theoretical Algorithm Design} Radio localization algorithms can be broadly categorized into two types: (i) two-stage localization and (ii) direct localization. Two-stage methods first estimate channel parameters, such as the \ac{aod}, and then infer the user’s location using geometric relationships. In contrast, direct localization approaches, such as electromagnetic fingerprinting, rely on databases of pre-recorded signal characteristics and match real-time signals to determine the user’s position. The introduction of RISs in radio localization has heightened the need for advanced estimation algorithms. For example, the inclusion of RISs introduces additional channel parameters in the two-stage localization scheme, necessitating improved channel estimation techniques. Several techniques, including atomic norm minimization, tensor decomposition, and deep learning, have been actively investigated to address this challenge.

\subsubsection{Realistic Studies} Although the theoretical algorithm development has been extensively studied for RIS-aided localization, experimental validation remains limited. These challenges arise from the high fabrication costs of high-frequency, wideband RISs, as well as from the mismatch between actual hardware characteristics and theoretical models. Hardware impairments, such as geometric misalignment, amplitude-phase dependencies, and element mutual coupling, often lead to the failure of theoretical localization algorithms in real-world scenarios~\cite{Ghazalian2024Calibration}. Therefore, a deep understanding of these non-ideal characteristics is essential before making the technology of RIS-aided localization a reality.

\subsection{How Can RIS Enhance SIOL?}

\begin{figure*}[t]
    \centering
    \begin{tikzpicture}
    \node (image) [anchor=south west]{\includegraphics[width=0.95\linewidth]{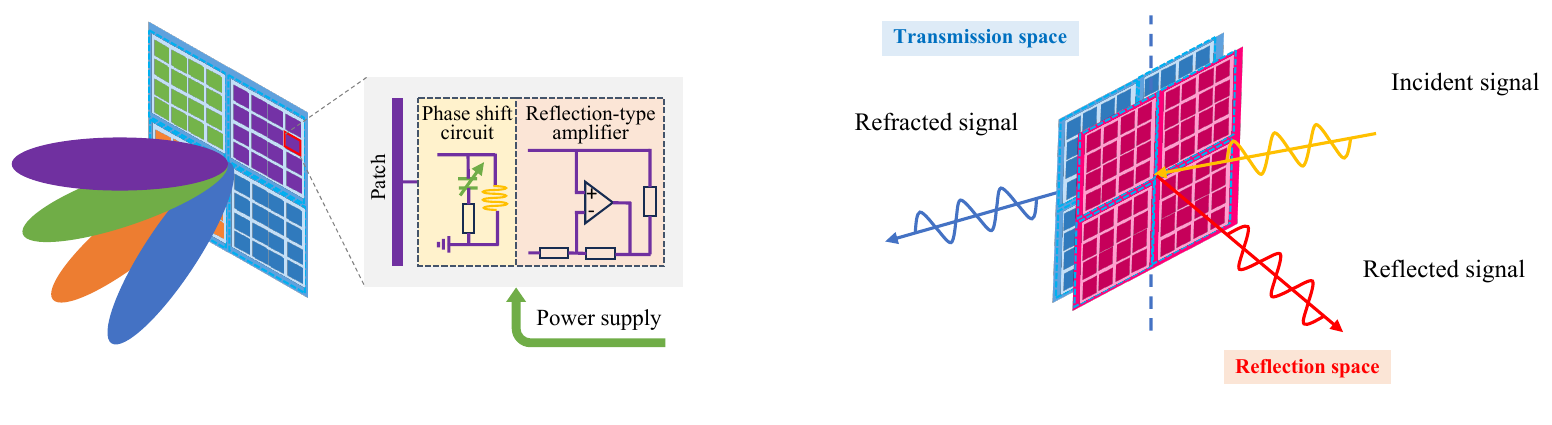}};
    \gettikzxy{(image.north east)}{\ix}{\iy};
    \node at (0.25*\ix,0.1*\iy)[rotate=0,anchor=north]{\footnotesize{(a) Active RIS}};
    \node at (0.74*\ix,0.1*\iy)[rotate=0,anchor=north]{\footnotesize{(b) STAR-RIS}};
    \node at (0.903*\ix,0.75*\iy)[rotate=0,anchor=north]{\footnotesize{$P_\text{in}$}};
    \node at (0.925*\ix,0.34*\iy)[rotate=0,anchor=north]{\footnotesize{$P_\text{reflected}=\epsilon^2P_\text{in}$}};
        \node at (0.59*\ix,0.67*\iy)[rotate=0,anchor=north]{\footnotesize{$P_\text{refracted}=(1-\epsilon^2)P_\text{in}$}};
	\end{tikzpicture}
    	\caption{Active RIS with reflection-type amplifiers and STAR-RIS with independent control of refraction (i.e., transmission) and reflection.}
    \label{fig:RIS}
\end{figure*}

To unveil the role that RISs can play in \ac{siol}, this paper places particular emphasis on two types of RIS: the active RIS and the simultaneous transmission and reflection RIS (STAR-RIS). These two types take center stage due to their distinctive attributes that can be leveraged to effectively strengthen the connection between LEO satellites and indoor users and thus enhance \ac{siol}. The conceptual representations of the active RIS and the STAR-RIS are shown in Fig.~\ref{fig:RIS}.

\subsubsection{Active RIS}
As illustrated in Fig.\ref{fig:RIS}-(a), the integration of active reflection-type amplifiers enables active RISs to extend their functionality beyond merely shifting the phase of incident signals to include signal amplification. This capability mitigates the so-called ``multiplicative fading'' effect and further enhances signal reception~\cite{Zhang2023Active}. For LEO-based \ac{siol}, active RISs can significantly improve localization performance by amplifying weak satellite signals, which is particularly critical for indoor users. 

\subsubsection{STAR-RIS}
Since the reflecting-only RIS only has a half-space coverage, the STAR-RIS has been proposed to obtain a full-space coverage~\cite{Ahmed2023A}, as depicted in Fig.~\ref{fig:RIS}-(b). The STAR-RIS is integrated with multiple functionalities, e.g., refraction (a.k.a. transmission) and reflection, which can provide a bridge between indoor and outdoor connectivity. In this sense, an outdoor BS (from either terrestrial or non-terrestrial networks) can offer extended coverage for indoor users while maintaining outdoor coverage with the aid of a STAR-RIS~\cite{Ahmed2023A}. By following this principle, the STAR-RIS can be a technological enabler for \ac{siol}. With the introduction of energy splitting between the two different modes, the quality of service (QoS) of the indoor and outdoor users can be concurrently adjusted.

\section{Simulation Studies}

This section conducts two simulation studies to demonstrate the potential of LEO satellite and RIS collaboration for \ac{siol} applications. To assess the achievable performance of different systems, this section focuses on evaluating the theoretical \ac{peb} derived from the \ac{crlb} theory, excluding considerations of practical algorithm implementation.

\begin{figure}[t]
  \begin{minipage}[b]{1\linewidth}
    \centering
      \include{figures/sim01-v2.0}
      \vspace{-3em}
  \end{minipage}
  \caption{
  Theoretical PEB vs. the number of LEO/MEO satellites, comparing scenarios with and without active STAR-RIS. 
  }
  \label{fig_sim01}
\end{figure}

\begin{figure}[t]
  \centering
  \includegraphics[width=3.5in]{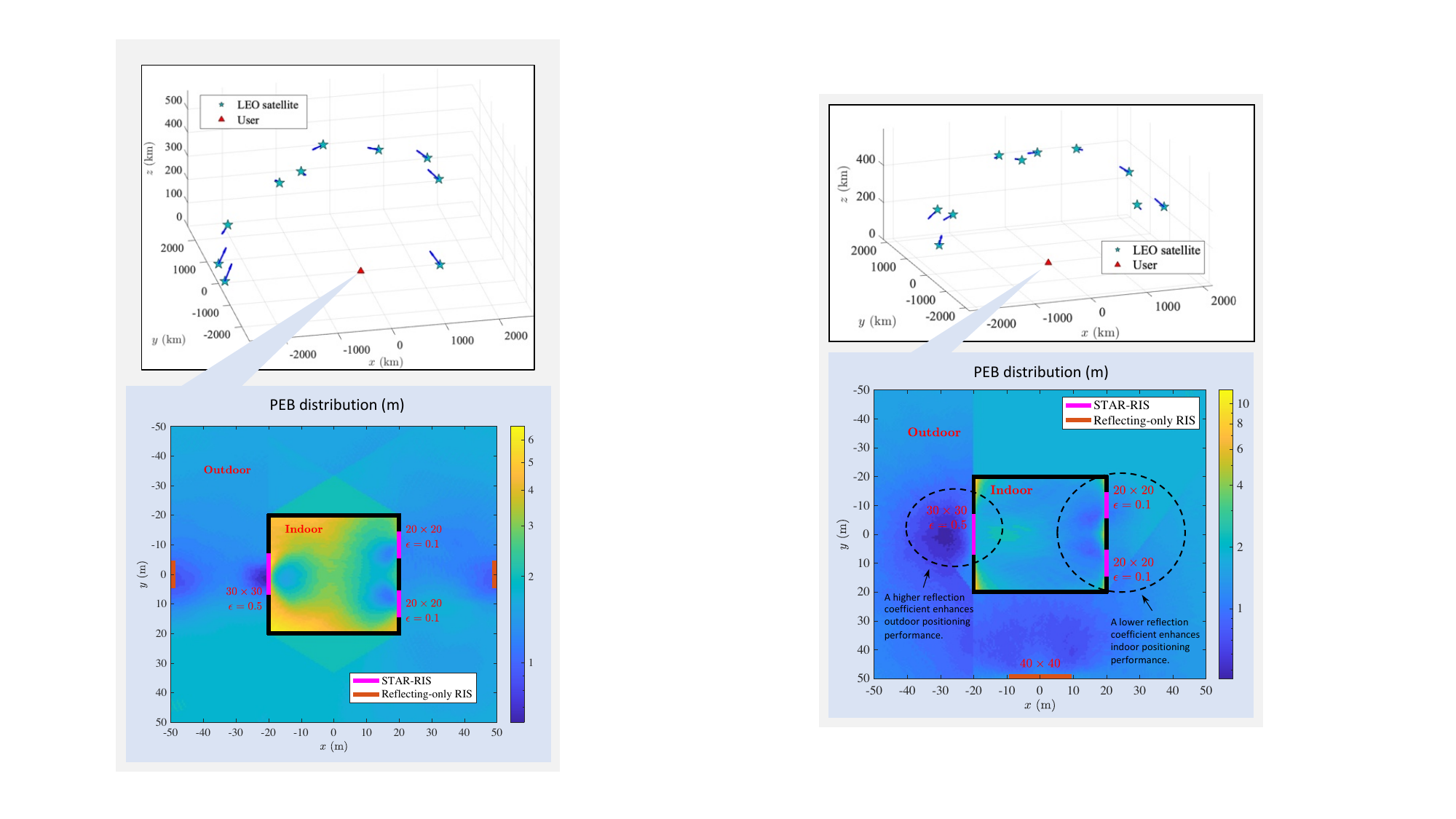}
  \caption{ 
      PEB evaluation (presented in units of meters) of simultaneous indoor and outdoor localization within a $\unit[100]{m}\times\unit[100]{m}$ region. Here, 10 randomly selected LEO satellites are used and presented from the user's perspective. These blue arrows visualize the velocities of the LEO satellites.
    }
  \label{fig_sim02}
\end{figure}

\subsubsection{Simulation Setup}
In our simulations, LEO satellites are located at an altitude of \unit[600]{km} with a carrier frequency of \unit[28]{GHz}, while MEO satellites are located at \unit[10,000]{km} with a carrier frequency of \unit[1.575]{GHz}, which are typical setup adopted in 3GPP TR 36.763 as presented in Fig.~\ref{fig_LinkBudget}. The satellite orbits are randomly generated using the QuaDRiGa simulator~\cite{Burkhardt2014QuaDRiGa}, and each satellite is equipped with an $8\times 8$ array of isotropic antennas spaced at half-wavelength intervals. Without assuming channel state information, both the LEO satellite antenna array and the RIS employ random beamforming designs, with the transmit power of each satellite set to \unit[55]{dBm} and the power supply of each RIS set to \unit[0]{dBm}. Downlink transmissions utilize OFDM signals with 3,000 subcarriers across a bandwidth of \unit[300]{MHz}, and the derived \ac{peb} evaluates the \ac{crlb} for estimating the user's position based on signals received over 5 transmissions. This performance is assessed based on the Fisher information in available channel parameters including AoDs from LEO satellites and RISs, time delays from LEO and RIS signals, and Doppler frequency shifts from LoS signals between LEO satellites and the user.

\subsubsection{PEB vs. Number of Satellites}

Fig.~\ref{fig_sim01} presents the PEB evaluation vs. the number of LEO and MEO satellites for both indoor and outdoor scenarios. An active STAR-RISs (a joint design of active RIS and STAR-RIS from Fig.~\ref{fig:RIS}), featuring a $20\times 20$ element configuration, is employed. We set the reflection coefficient $\epsilon=0.5$, indicating the proportion of power reflected as illustrated in Fig.~\ref{fig:RIS}. Throughout all the tested scenarios, the distance between each user and the STAR-RIS is fixed as \unit[15]{m}.

As shown in Fig.~\ref{fig_sim01}, increasing the number of LEO satellites generally leads to a reduction in PEB, reflecting improved localization accuracy. However, performance saturation is observed when the number of satellites exceeds 6. Comparing the LEO evaluations with and without RIS (depicted by the blue and red curves), we find that incorporating active STAR-RIS results in a significant reduction in PEB, exceeding an order of magnitude. Additionally, by comparing LEO and MEO satellites (depicted by the red and green curves), it is evident that LEO satellites improve PEB by approximately an order of magnitude relative to MEO satellites for both indoor and outdoor users. Notably, for the most challenging indoor localization scenarios, only the integration of LEO satellites with active STAR-RIS successfully achieves the theoretical meter-level accuracy, highlighting the potential of these two technologies for \ac{siol}.

\subsubsection{PEB vs. User's Positions}

Now we evaluate the localization performance under varying user positions. Fig.~\ref{fig_sim02} provides a graphical representation of the 10 satellites employed in this simulation, while the zoomed-in subfigure illustrates the evaluation of the \acp{peb} across diverse user locations, encompassing both indoor and outdoor environments within a $\unit[100]{m}\times\unit[100]{m}$ region. It is evident that with the involvement of multiple satellites and RISs, it is possible to achieve meter-level localization accuracy consistently. Notably, active STAR-RISs hold the potential to enable indoor localization. The indoor regions in close proximity to STAR-RISs can achieve even more accurate localization surpassing the precision of certain outdoor areas. Furthermore, the results in Fig.~\ref{fig_sim02} demonstrate the impact of the STAR-RIS reflection coefficient~$\epsilon$. Specifically, higher values of $\epsilon$ enhance outdoor localization performance, while lower values improve indoor localization accuracy. This suggests that by adjusting the reflection coefficient of STAR-RISs, it is possible to balance indoor and outdoor localization performance to meet different service requirements. 

\section{Architecture Discussion and Key Challenges}

\begin{figure*}[t]
  \centering
  \includegraphics[width=\linewidth]{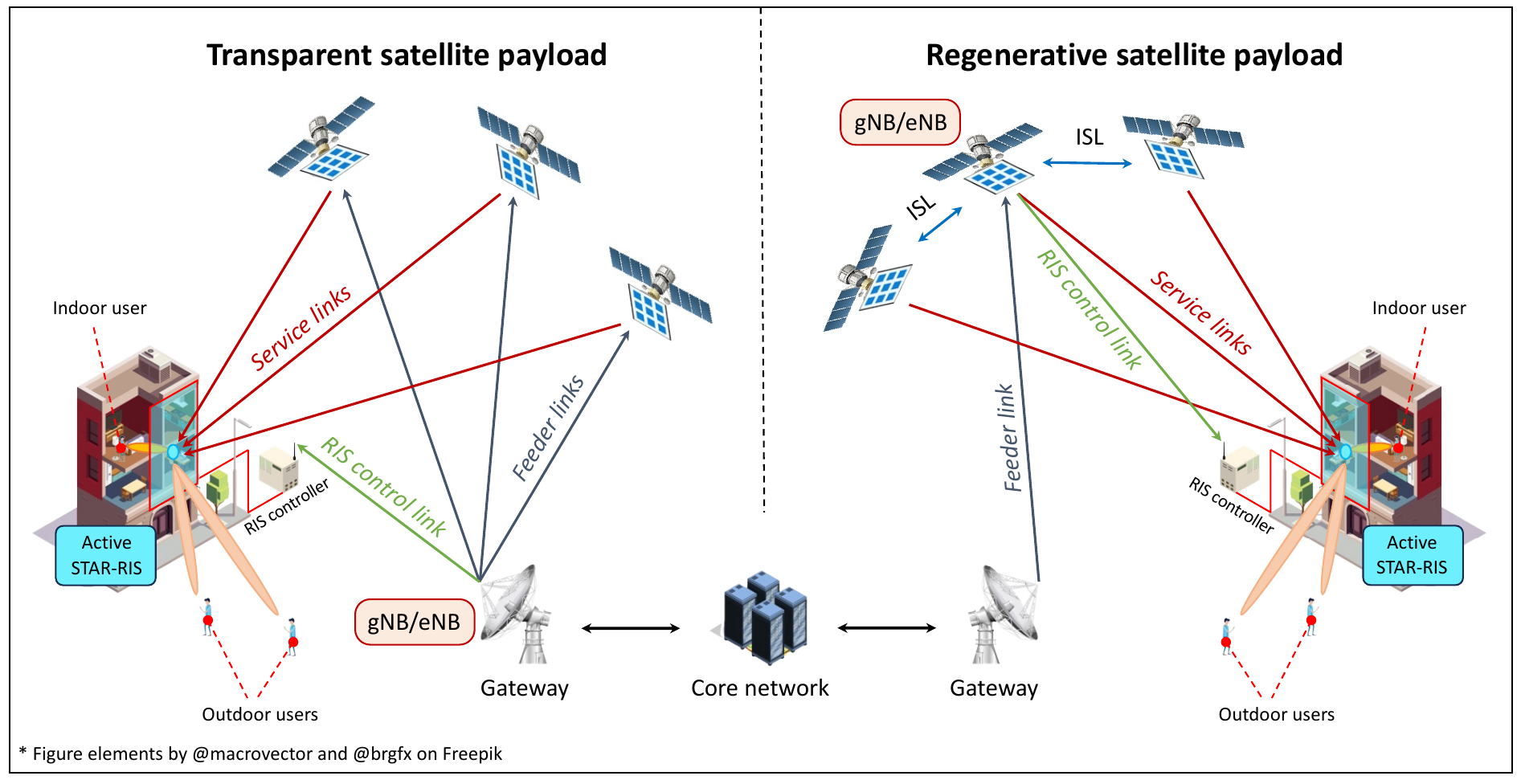}
  \caption{ 
      Possible system operation architectures for coordinating LEO satellite and RIS based on transparent and regenerative satellite payloads. The satellite feeder links and RIS control links are configured to connect gNB/eNB, while gNB/eNB can be situated at a terrestrial gateway or on a designated central satellite.
    }
  \label{fig_architecture}
\end{figure*}

While the numerical analysis highlights the significant potential of LEO satellite and RIS collaboration for SIOL applications, practical implementation requires careful consideration of system architectures. In this section, we first explore practical system architectures and then outline several key challenges in implementing such systems.

\subsection{System Architecture}

Depending on whether including signal transformation functionality, LEO satellites may implement either a transparent or a regenerative (with onboard processing) payload~\cite{3GPP38821}. Fig.~\ref{fig_architecture} illustrates the possible \ac{siol} system architectures for both transparent and regenerative satellite payload scenarios.
\begin{itemize}
    \item \textbf{Transparent Satellite Payload:} In this configuration, satellites act solely as relays, transmitting signals to ground users without any processing capabilities. The terrestrial gateway functions as the gNB/eNB, performing two key tasks: (i) transmitting data to the satellites via feeder links and (ii) managing the RIS response through a dedicated RIS control link.
    \item \textbf{Regenerative Satellite Payload:} In this configuration, the satellite incorporates gNB/eNB functionality, enabling onboard processing. A central satellite is designated to receive data from the terrestrial gateway and distribute it to other satellites via inter-satellite links (ISLs). Simultaneously, the central satellite can control the RIS response through a satellite-to-RIS control link.
\end{itemize}

Both architectures encounter numerous common challenges in practice that have not been adequately addressed and require more effective solutions, as discussed in the following subsection.

\subsection{Key Challenges}

\subsubsection{Signal Processing Challenges}
The integration of LEO satellites and RIS presents significant challenges for estimation algorithms. On the one hand, both LEO satellites and RIS introduce more potential observations than conventional radio localization systems (such as GNSS), requiring efficient algorithms for complex data fusion. On the other hand, RIS introduces extra reflection paths, which demand advanced algorithms and codebook designs to separate and extract information in various paths from the received signals. In addition to localization algorithms, joint beamforming design for satellites and RIS also remains a challenge, particularly when considering the large and varying propagation delays in satellite-to-RIS links. Precise control of the RIS response to reflect LEO signals at the correct time is essential. Accurate timing advance estimation is critical to addressing this issue.

\subsubsection{Handover Management}
Due to the high mobility of LEO satellites, the handover process between satellites occurs frequently. This may result in significant signaling overhead and service interruption, affecting both localization and communication services. Several triggering methods, such as measurement-based, location-based, time-based, and timing advance value-based triggering, have been studied in 3GPP TR 38.821~\cite{3GPP38821} to mitigate the handover rate in various scenarios. In addition to optimizing system handover management, integrating LEO satellites with other local non-terrestrial networks, such as high-altitude platform stations (HAPS) and unmanned aerial vehicles (UAVs), offers an alternative solution to ensure uninterrupted services.

\subsubsection{STAR-RIS Hardware Fabrication}
Although STAR-RISs have shown significant promise for achieving so-called full-space manipulation of signal propagation, and several theoretical performance assessment works have been reported~\cite{Ahmed2023A}, the hardware realization of STAR-RISs remains blank in the literature. The prospects for the hardware realization of STAR-RIS may stem from the integration of the transmitarray and reflectarray antennas. However, challenges may arise in the effective control of transmitting and reflecting power allocation, the independent reconfiguration of transmitting and reflecting phase shifts, and the mitigation of interference within the feeding network.

\section{Conclusion}

This article proposes a potential solution for achieving global, seamless indoor and outdoor localization by leveraging the cooperation between LEO satellites and RISs. The feasibility and suitability of integrating these two technologies for uninterrupted localization services are demonstrated through comprehensive analyses and theoretical simulations. The discussion on system architecture and key unresolved challenges within this domain underscores the importance of ongoing exploration in this evolving field.

\section{Acknowledgments}
Fig.~\ref{fig_system} was created by Thom Leach, a Scientific Illustrator at KAUST.

\bibliography{references}
\bibliographystyle{IEEEtran}

\vspace{-1cm}
\begin{IEEEbiographynophoto}{Pinjun~Zheng}
(pinjun.zheng@kaust.edu.sa) is a PhD student in the Electrical and Computer Engineering Program, CEMSE, King Abdullah University of Science and Technology (KAUST), Thuwal, 23955-6900, Kingdom of Saudi Arabia. His research interest is signal processing for localization and communication. \end{IEEEbiographynophoto}

\vspace{-1cm}
\begin{IEEEbiographynophoto}{Xing~Liu}
(xing.liu@kaust.edu.sa) is a postdoctoral researcher in the Electrical and Computer Engineering Program, CEMSE, King Abdullah University of Science and Technology (KAUST), Thuwal, 23955-6900, Kingdom of Saudi Arabia. His current research focuses on precise GNSS positioning and
attitude determination.
\end{IEEEbiographynophoto}

\vspace{-1cm}
\begin{IEEEbiographynophoto}{Yuchen~Zhang}
(yuchen.zhang@kaust.edu.sa) is a postdoctoral researcher in the Electrical and Computer Engineering Program, CEMSE, King Abdullah University of Science and Technology (KAUST), Thuwal, 23955-6900, Kingdom of Saudi Arabia. His current research focuses on NTN communication and positioning and ISAC.
\end{IEEEbiographynophoto}

\vspace{-1cm}
\begin{IEEEbiographynophoto}{Jiguang~He}
(jiguang.he@tii.ae) is a senior researcher at the Technology Innovation Institute, Masdar City, United Arab Emirates. His research interests span millimeter wave MIMO communications, reconfigurable intelligent surfaces for simultaneous localization and communication (SLAC), and advanced signal processing techniques.
\end{IEEEbiographynophoto}

\vspace{-0.5cm}
\begin{IEEEbiographynophoto}{Gonzalo~Seco-Granados}
(gonzalo.seco@uab.cat) is a professor with the Department of Telecommunication, Universitat Autònoma de Barcelona, Bellaterra, Spain. His research interests include GNSS, and beyond 5G integrated communications, localization, and sensing.
\end{IEEEbiographynophoto}

\vspace{-0.5cm}
\begin{IEEEbiographynophoto}{Tareq~Y.~Al-Naffouri}
(tareq.alnaffouri@kaust.edu.sa) is a professor in the Electrical and Computer Engineering Program, CEMSE, King Abdullah University of Science and Technology (KAUST), Thuwal, 23955-6900, Kingdom of Saudi Arabia. His current research focuses on the areas of sparse, adaptive, and statistical signal processing and their applications, as well as machine learning and network information theory.
\end{IEEEbiographynophoto}

\vfill

\end{document}

%% file: figures/Spider_Chart.tex
\usepgfplotslibrary{polar}

\begin{tikzpicture}
\begin{polaraxis}[
    width=7.5cm,
    height=5cm,
    ymin=0,
    ymax=10,
    yticklabels={},
    xtick={0,45,...,315},
    xticklabels={Lifespan, Latency, , Orbit Altitude, Data Throughput, , , Launch Cost},
    rotate=22.5, 
    axis on top,
    grid=both,
    minor tick num=1,
    major grid style={white},
    minor grid style={gray!25},
    legend style={at={(-0.05,1.48)}, anchor=north west, font=\fontsize{7}{8.}\selectfont},
    font=\small
]

\addplot+[mark=none, ultra thick, draw=green, fill=green, fill opacity=0.3] coordinates {
    (0,10)
    (45,8)
    (90,10)
    (135,9)
    (180,6)
    (225,2)
    (270,5)
    (315,10)
    (0,10) 
};

\addplot+[mark=none, ultra thick, draw=blue, fill=blue, fill opacity=0.3] coordinates {
    (0,7)
    (45,6)
    (90,7)
    (135,6)
    (180,7)
    (225,5)
    (270,6)
    (315,7)
    (0,7) 
};

\addplot+[mark=none, ultra thick, draw=red, fill=red, fill opacity=0.3] coordinates {
    (0,5)
    (45,2)
    (90,4)
    (135,2)
    (180,9)
    (225,9)
    (270,9)
    (315,4)
    (0,5) 
};

\legend{GEO, MEO, LEO}

\end{polaraxis}

\coordinate (A) at ([shift={(85:-2.25cm)}]current axis.origin);
\node[anchor=east] at (A) {\small Maneuverability};
\coordinate (B) at ([shift={(-30:3.95cm)}]current axis.origin);
\node[anchor=east] at (B) {\small Operational Complexity};
\coordinate (C) at ([shift={(90:2.00cm)}]current axis.origin);
\node[anchor=east] at (C) {\small Coverage Area};

\end{tikzpicture}

%% file: figures/sim01-v2.0.tex
\definecolor{ForestGreen}{rgb}{0.1333    0.5451    0.1333}
\definecolor{DarkGoldenrod}{rgb}{0.9333    0.6784    0.0549}
\definecolor{BlueViolet}{rgb}{ 0.6275    0.1255    0.9412}
\definecolor{Firebrick}{rgb}{0.8039    0.1490    0.1490}
\definecolor{mycolor1}{rgb}{1.00000,0.00000,1.00000}%
\begin{tikzpicture}

\begin{axis}[%
  width=2.8in,
  height=1.9in,
  at={(0in,0in)},
scale only axis,
xmin=1.93,
xmax=8.07,
xlabel style={font=\color{white!15!black},font=\footnotesize},
xticklabel style = {font=\color{white!15!black},font=\footnotesize},
xlabel={Number of satellites},
ymode=log,
ymin=0.07,
ymax=100,
yminorticks=true,
ylabel style={font=\color{white!15!black},font=\footnotesize},
yticklabel style = {font=\color{white!15!black},font=\footnotesize},
ylabel={PEB (m)},
axis background/.style={fill=white},
xmajorgrids,
ymajorgrids,
yminorgrids,
legend style={at={(1,1.05)}, anchor=south east, legend columns=2, legend cell align=left, align=left, draw=white!15!black, font=\scriptsize}
]
\addplot [color=blue, mark=triangle, line width=1pt, mark options={solid, blue}, mark size=3pt]
  table[row sep=crcr]{%
2	21032087.7450877\\
4	289.210966462205\\
6	59.5345868664781\\
8	44.4992336672473\\
};
\addlegendentry{LEO, indoor, without RIS}

\addplot [color=blue, mark=triangle, line width=1pt, mark options={solid, rotate=180, blue}, mark size=3pt]
  table[row sep=crcr]{%
2	805599.226268519\\
4	11.0651531008214\\
6	2.27784887623631\\
8	1.70262705853685\\
};
\addlegendentry{LEO, outdoor, without RIS}

\addplot [color=ForestGreen, dashed, mark=triangle*, line width=1.3pt, mark options={solid, ForestGreen}, mark size=2.5pt]
  table[row sep=crcr]{%
2	68.5715276811637\\
4	39.9933628815632\\
6	26.4211722909613\\
8	24.7531243994839\\
};
\addlegendentry{MEO, indoor, with RIS}

\addplot [color=ForestGreen, dashed, mark=triangle*, line width=1.3pt, mark options={solid, rotate=180, ForestGreen}, mark size=2.5pt]
  table[row sep=crcr]{%
2	2.82612804231038\\
4	1.1148111226968\\
6	0.973508380145195\\
8	0.925784913416708\\
};
\addlegendentry{MEO, outdoor, with RIS}

\addplot [color=red, mark=triangle*, line width=1pt, mark options={solid, red}, mark size=2.5pt]
  table[row sep=crcr]{%
2	36.7414545457404\\
4	1.80899913697245\\
6	1.41031722865898\\
8	1.40993791507766\\
};
\addlegendentry{LEO, indoor, with RIS}

\addplot [color=red, mark=triangle*, line width=1pt, mark options={solid, rotate=180, red}, mark size=2.5pt]
  table[row sep=crcr]{%
2	0.893763247177418\\
4	0.11675248985969\\
6	0.0914670669382468\\
8	0.0896706480271687\\
};
\addlegendentry{LEO, outdoor, with RIS}

\end{axis}
\end{tikzpicture}%